\documentclass{article}

\usepackage{arxiv}

\usepackage[utf8]{inputenc} 
\usepackage[T1]{fontenc}    
\usepackage{hyperref}       
\usepackage{url}            
\usepackage{booktabs}       
\usepackage{amsfonts}       
\usepackage{nicefrac}       
\usepackage{microtype}      
\usepackage{lipsum}
\usepackage{graphicx}
\graphicspath{ {./images/} }

\usepackage{hyperref}
\usepackage{amsmath}
\usepackage{amssymb}
\usepackage{xcolor}
\usepackage{bm}
\usepackage{parskip} 
\usepackage{optidef}
\usepackage{multirow} 

\newcommand*{\Scale}[2][4]{\scalebox{#1}{$#2$}}%
\usepackage{graphicx}

\ifLuaTeX
  \usepackage{selnolig}  
\fi
\usepackage[]{natbib}
\usepackage{bookmark}

\usepackage{mathtools}
\mathtoolsset{showonlyrefs} 

\definecolor{main}{HTML}{5989cf}    
\definecolor{sub}{HTML}{cde4ff}     

\usepackage[many]{tcolorbox}    	

\usepackage{setspace}               
\usepackage{multicol}               

\newtcolorbox{boxC}{
    colback = sub, 
    boxrule = 0pt  
}

\title{\bf Explaining and Connecting Kriging with Gaussian Process Regression}

\author{
 Marius Marinescu \\
  Engineering School of Fuenlabrada\\
  King Juan Carlos University \\
  Madrid, Spain \\
  \texttt{marius.marinescu@urjc.es} \\
}
\begin{document}
\maketitle
\begin{abstract}
Kriging and Gaussian Process Regression are statistical methods that allow predicting and quantifying the uncertainty of random field outcomes by using a sample of correlated observations. The methods have different origins. Kriging comes from geostatistics, a field which started to develop around 1950 oriented toward mining valuation problems, whereas Gaussian Process Regression has gained popularity in the area of machine learning since the late 20th century. In the literature, the methods are often described as equivalent. However, beyond this assertion, thorough comparisons are notably absent. Furthermore, Kriging has many variants, and this statement should be clarified. In this paper, this gap is filled.
The three classical versions of Kriging are considered: Simple Kriging, Ordinary Kriging and Universal Kriging. 
%
It is shown that the methods are closely related, however, differ in their assumptions and statistical approach, much like the least squares method, the BLUE method, and the likelihood method in regression do. The study provides useful insights into the interplay between the methods and serves as a cohesive resource for researchers and practitioners entering the field, thereby facilitating the transfer of knowledge between them.
\end{abstract}

\textbf{Key words}. Kriging, Gaussian processes, Regression


\section{Introduction}\label{sec:intro}

In the literature about Kriging, is not uncommon to find statements suggesting that Kriging and Gaussian Process Regression (GPR) are the same. For example, in \citet[pg. 30]{rasmussen2006} we find `\textit{Gaussian process prediction is also well known in the geostatistics field, where it is known as kriging}' or in \citet[Chapter 5]{gramacy2020} when referring to GPR: `\textit{The subject of this chapter goes by many names and acronyms. Some call it kriging, which is a term that comes from geostatistics}'.

I found it notable that beyond these affirmations, there is no precise and rigorous derivation explaining why they are equal, and under which conditions. Furthermore, Kriging has many variants and this affirmation appears to be, at least, too vague. 
%
On the other hand, the literature about Kriging is huge and sometimes lacks enough thoroughness. Moreover, the consensus for the terminology and notation is weak, and it lacks of an adequate standardisation. This has already been noticed by some authors, for instance in \citet[Sec. 2]{YAKOWITZ1985} it is stated `\textit{There are some inconsistencies in the fundamental definitions and results in the kriging literature. For example, the definitions of “intrinsic random function” given by David 
[6] and Matheron [30] do not coincide. (...) This multiple use has not been carefully distinguished by kriging authors, and there has been resulting discrepancy in the mathematical 
representations.}'. In \citet[Preface]{stein1999} it is stated `\textit{Section 6.3, points out an important error in Matheron (1971) ...}', or in \citet[Sec. 29.8]{delfiner2012} which declares `\textit{We also gave a look at current research to enable a global application of kriging (...) Much work remains necessary to transform them in standard methods applicable to a large variety of situations}'. See also commentaries before Eq. 9, Sec. 2 in \citet{kleijnen2017} and \citet{demystifying2023}.

In light of these inconsistencies and to ensure clarity of exposition, the Kriging equations will be derived from first principles, which is also necessary for a thorough comparison with GPR. 


Thus, the main objective of this work is to present and clarify the connections between Kriging and GPR.
The work is mostly self-contained, aiming to be expository and clear, but maintaining the required rigour.
A secondary objective  is to facilitate the transfer of knowledge between fields (as these topics are typically scattered across unrelated literature).



The contributions of this paper are as follows: 1) a novel historical introduction of Kriging using modern statistical terminology, 2) a variant-by-variant comparison of Kriging and GPR, 3) a joint and unified mathematical presentation, and 4) a statistical analysis establishing their connections (including similarities and differences).

The remainder of this paper\footnote{A poster version of this work was presented at the Spatial Statistics 2025: At the Dawn of AI congress and is available at \citet{marinescu2024poster}.} is organized as follows. The rest of Section~\ref{sec:intro} gives historical background. Section~\ref{sec:2} provides a comprehensive derivation of classical Kriging. Section~\ref{sec:stairs} bridges the methodologies by embedding Kriging within the linear regression framework, establishing a common baseline for the comparison with GPR. Section~\ref{sec:gpr} presents the mathematical formulation of GPR. Section~\ref{sec:5} analyses their connections. Finally, Section~\ref{sec:concl} concludes the paper.


\subsection{Kriging historical introduction}


    

Daniel Gerhardus Krige worked in the gold mines of the Witwatersrand Basin, in South Africa \citep{krige2015}. One of the primary challenges in these mines was to accurately estimate the gold content of ore bodies. 

From a techno-economic perspective, a mine deserves to be exploited only if the cost of its extraction and processing does not exceed the value of the metal which can be extracted from it. The true grade, i.e. the amount of valuable material per unit of rock, of a panel\footnote{In the context of mining, a panel refers to a specific section or subdivision of a mine that is being worked on or has been prepared for extraction.} is not known before
its exploitation, so it is estimated by using a sample. At the beginning of the 1950s, the estimate was simply the average grade of the data
belonging to the panel or situated at its border. D. Krige noticed that gold deposits exhibited spatial continuity and that observations were not independent but correlated, and was struck by the fact that, on average, low-grade panels were underestimated and high-grade panels were overestimated. 

In statistical terms, suppose $Z_p$ is the random variable representing the average grade of a panel. Suppose that $\mathbb{E}[Z_p]=m$ holds, where $m$ is the average amount of gold in the whole mine. Then, what D. Krige noticed is:
%
%
\begin{align}
    m &< \mathbb{E}[Z_p \ | \  \bar{Z}=\bar{z}, \ \bar{z}>m]< \bar{z} \label{eq:under} \\
    \bar{z} &< \mathbb{E}[Z_p \ | \  \bar{Z}=\bar{z}, \ \bar{z}<m]< m \label{eq:over}
\end{align}


%
%
where $\bar{Z}$ is the sample average of a panel. 
%
D. Krige was not the first to notice this under/over valuation of panel grades, but was the first to convincingly use statistical theory to tackle this problem \citep{krige1951, krige1962}.

\begin{figure}
    \centering
    \includegraphics[width=.9\linewidth]{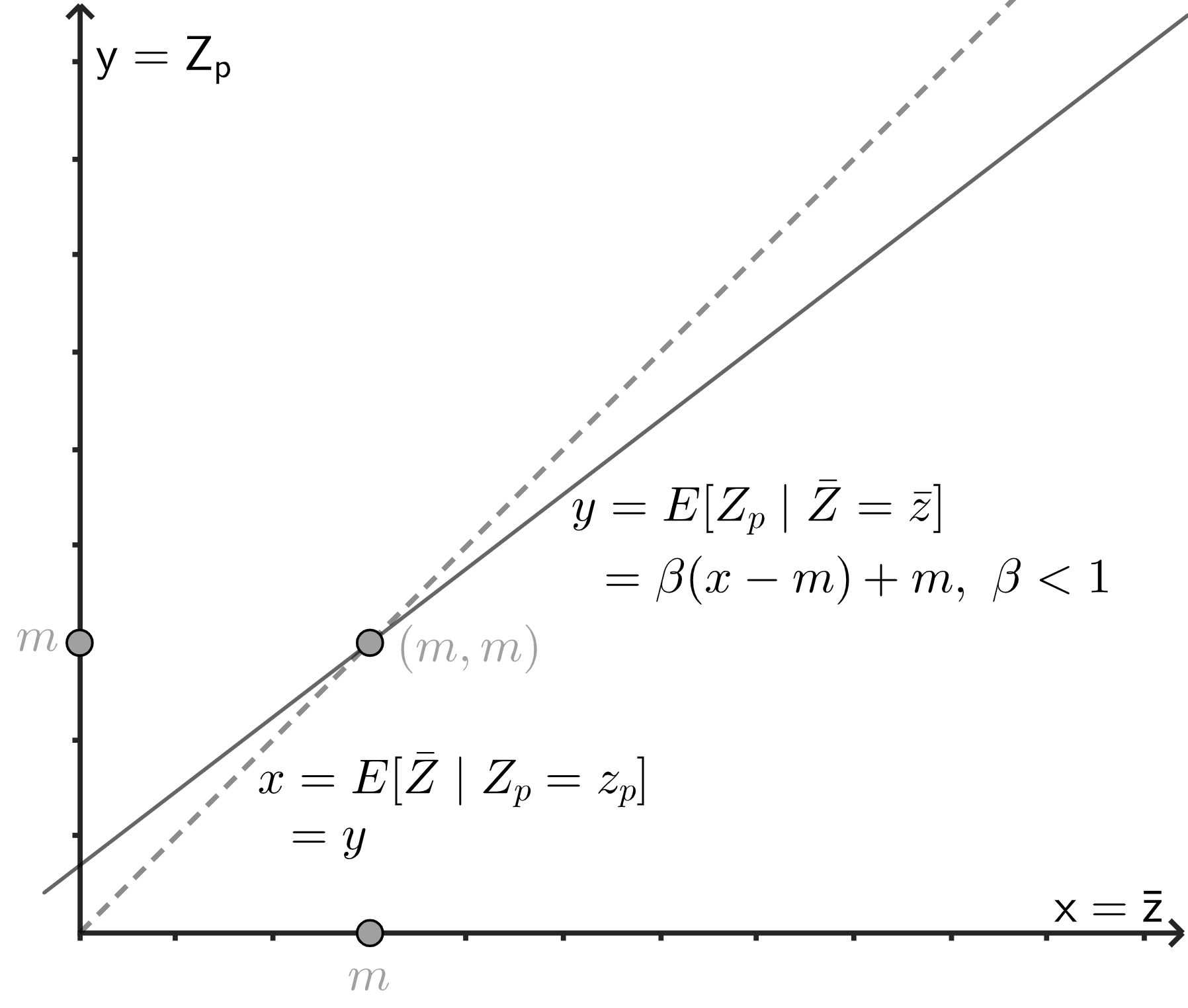}
    \caption{D. Krige, observing the data, hypothesized that $E[\bar{Z} \mid Z_p=z_p]=z_p$ holds (dotted line), but the  equality in the converse conditional expectation does not hold (continuous line). 
    The observed under/over valuation of panels, reflected in Eqs. \eqref{eq:under} and \eqref{eq:over}, implies that $1=\rho\frac{\sigma_\text{\tiny X}^2}{\sigma_Y^2}$ which makes $\beta = \frac{\sigma_\text{\tiny Y}^2}{\sigma_X^2}$, and that $\beta <1$.}
    %
    \label{fig:reg_line}
\end{figure}

He observed that panel grade observations came close to the log-normal probability law (e.g. see \citet[Diagram 2]{krige1951}), and applied classical regression theory, over the axles $y= Z_p, \ x=\bar{z}$ (e.g. see \citet[Diagram 24 and 25]{krige1962}). 
%
A schematic representation of the regression line is shown in Fig. \ref{fig:reg_line}, to which we will come back later on.

By using his method, the systematic error of using the sample average was tackled from an appropriate statistical perspective and the improvement of the accuracy in estimating the panel grades was high, see \citet[Sec. 6 - Improved Estimates Based on Statistical Theory]{krige1951}.

One of the reasons that motivated D. Krige work was the rudimentary use of statistics in mining at that time. I cite two of his statements from \citet{krige1951}:
``\textit{At present these methods consist almost 
entirely of the application of simple arithmetic and empirical formulae guided by practical experience and ignore the many advantage to be gained from a carefully statistical analysis ...}' and
`\textit{Even an experienced mine valuator on 
the Rand may believe that the variation 
between gold values along a stretch of drive, raise or stope face is haphazard. This is not the case, (...)}'.
I refer to the previous papers for a more detailed view of D. Krige's concerns.

On the other hand, the (spatial) auto-correlation structure between observations was not taken into account, making estimators non-efficient. Here is where G. Matheron came into. 
%
The related French community started to use routinely the term `Le Krieage'  and G. Matheron coined the term `Kriging' in his publications in honour of D. Krige's work \citep{cressie1990}. 
He urged all scientists concerned with spatial interpolation to adopt this term and afterward became common in the Anglo-Saxon mining terminology.
What Kriging did and `Kriging' as coined by Matheron is not the same idea. 
Given a random field, Matheron defined Kriging as a way to predict an unobserved value or a block average using the available observations. In particular, but without using this term, he derived the Best Linear Unbiased Predictor (BLUP) in the spatial statistics setting \citep{matheron1971}.

The connection between them can be argued as follows.
The regression line used by D. Krige was of the form \citep[Eq. 3-1]{matheron1971}: 
%
\begin{equation}
    y=m+\beta (x-m), \ \ \beta < 1.
\end{equation}
%


In Fig. \ref{fig:reg_line} the regression line is represented. 
Both $m$ and $\beta$ have to be estimated from a large enough collection of data from several panels. D. Krige estimated it by using classical regression theory \citep[Eq. 22]{krige1962}. The key is that for estimating $m$ he used the ordinary estimator $\hat{m}=\bar{X}=\frac{1}{n}\sum_{i=1}^n X_i$ which gives:
%
\begin{equation}\label{eq:mkri}
    y=\bar{X}+\beta (x-\bar{X})=\sum_{i=1}^n a_i X_i + \beta x
\end{equation}


%
with $a_i=\frac{1-\beta}{n}, \ i=1,...n$, and with $n$ being all the available data from the mine. Matheron argued that in spite of assigning a constant weight $a_i$ to each observation, appropriate weights, taking into account the location of the observation with respect to the panel, should be used. 
This actually results in what is known in modern statistics as the Best Linear Unbiased Estimator (BLUE), in this case particularised to geostatistics, where the covariances are extracted from the autocovariance function. Then Matheron applied the same logic, of using a linear combination of observations, to the problem of predicting an unobserved value $Z_*$ and coined it as Kriging.

It is worthwhile to come back to D. Krige's work and rewrite slightly Eq. \eqref{eq:mkri}  to observe another interesting perspective:
%
\begin{equation*}
    y=\bar{X}+\beta (x-\bar{X})=(1-\beta) \bar{X} + \beta x,  \ \  \beta < 1.
\end{equation*}


%
We can see that D. Krige estimated a panel's true grade by a convex sum of the sample average of the whole data on the mine and the sample average of the panel. Thus, he used a weighted sum, not of all observations, but of the global and local sample average.








Matheron is considered the creator of Geostatistics \citep{chiles2018}. In \citet{agterberg2004} he is presented as one of the greatest mathematical-statisticians from the twentieth century, at the standing of Ronald Fisher or John Tukey.

\subsection{GPR historical introduction}

The formalisation and widespread use of GPR as we know it today, have different origins, and began in the 1990s, largely driven by advancements in the machine learning field. 

The concept of GPR is based on the concept of Gaussian Process (GP).  One of the earliest contributors to the theoretical foundations of GPs was Andrey Kolmogorov who laid the groundwork for the theory of stochastic processes \citep{kolmogorov1938}. As a difference with Kriging whose development is more diffuse, Kolmogorov’s work on probability theory and stochastic processes provided a rigorous mathematical framework that would later be fundamental to the development of GPR. A seminal  book about the mathematical foundation of GP, written by Kolmogorov's former students, is \citet{ibragimov1978}.


The use of GP in estimation emerged from the works of pioneering statisticians and mathematicians who sought to understand and model random processes over time and space.  In the latter half of the 20th century, researchers recognised the versatility of GPs in modelling different type of data. This period saw the development of key theoretical advancements and practical applications. 
The late 1990s and early 2000s marked a significant turning point for GPR, driven by the rise of machine learning and the increasing availability of computational resources.  

Christopher Williams and Carl Rasmussen were pivotal figures in the transition to machine learning. Their influential textbook called Gaussian Processes for Machine Learning  \citep{rasmussen2006}, synthesised previous theoretical developments and applied them to a wide range of problems, including regression and classification. They demonstrated the utility of GPR in providing not just predictions but also measures of uncertainty through the posterior distribution. Also,  they have shown its connection to neural network models, among other types of models \citep[Chapter 6 and 7]{rasmussen2006}. 



GPR has some advantages, which made it gain popularity:
\begin{itemize}

    \item A clear mathematical foundation, based on stochastic processes.
    
    \item The capacity to provide a measure of uncertainty on its predictions.
    \item The flexibility of the model. By choosing and combining different kernels, GPR can model a wide range of target functions and capture different  data patterns.
    \item Its non parametric nature. Unlike parametric models, GPR does not assume a fixed form for the underlying (prior) functions, making it highly adaptable to complex datasets.
    \item GPR interpretability. The probabilistic nature and the explicit form of the covariance function make GPR interpretable, in opposition to black-box models.
\end{itemize}
In conclusion, the history of GPR is a testament of the interdisciplinary nature of modern statistical and machine learning methodologies. From its origins in the theoretical work of the early 20th century to its practical applications and its evolution into a fundamental tool in modern machine learning, GPR has continually adapted and expanded its scope. Today, it stands as a robust and versatile method for modelling complex data.

\section{Mathematical formulation of Kriging}\label{sec:2}

In this work the same notation convention as in \citet[Symbols and Notation]{rasmussen2006} is used. Vectors are represented in bold type, whereas matrices are capitalised. Random variables will also be capitalised. Estimators will be indicated with a hat, and when necessary with a superscript such as GLS, standing for Generalised Least Squares. Other expressions and abbreviations will be defined on scratch.
In the supplementary material \ref{ap:k_terminology}, some typical Kriging terminology used in the literature is explained. This terminology will not be used extensively in this paper since is not necessary, but you may find them useful for consulting and comparing the literature. 

Kriging observations are modelled as coming from a scalar-valued\footnote{Co-Kriging, which involves multiple outputs, is beyond the scope of this work.} random field which is a way to refer to stochastic processes indexed by a multi-set.

Formally, let $Z : D \times \Omega 	\longmapsto \mathbb{R} $ be a function of two arguments, $\bm{x} \in D$ and $\omega \in \Omega$, 
where $D$ is a subset of  $\mathbb{R}^d$ and  $(\Omega ,{\mathcal {F}},P) $ denotes a probability space. 
If $Z(\bm{x})$ is a random variable on the probability space $(\Omega ,{\mathcal {F}},P) $ for each $\bm{x} \in D$, then $Z$ is said to be an \textbf{scalar-valued random field} \citep[Sec 3.2]{grigoriu2013}. 




The main variants of Kriging are distinguished according to the trend model. This will give rise to three well established types of Kriging in the literature, namely: Simple Kriging (SP), Ordinary Kriging (OK) and Universal Kriging (UK), to which we will return at length. 

%

Let's consider that we have some observations from a random field $\{Z(\bm{x})\}_{\bm{x} \in \mathbb{R}^d }$, where the mean function $m(\bm{x})=E[Z(\bm{x})]$ and the autocovariance function $k(\bm{x}, \bm{x'})=$ \\ $\mathrm{Cov}(Z(\bm{x}), Z(\bm{x'}))$ exists, sometimes called a second order random field \citep{cressie2015}. Suppose that both are known. Consider also that we observe the random field with some (additive) noise: 
\begin{equation}
     Y(\bm{x}_i)=Z(\bm{x}_i)+\varepsilon_i, \ i=1, \ldots, n
\end{equation}

%

\noindent where $ \varepsilon_i $  are i.i.d. (independent and identically distributed) random variables. Suppose that we want to infer the value of the random field at a new position $\bm{x}_*$. As said, what Kriging pursues is to find what is called in statistics the Best Linear Unbiased Predictor\footnote{In the literature, the term BLUE is commonly used when estimating deterministic quantities, such as the parameters in classical regression. However, in this case, we estimate a random quantity, $Z(\bm{x}_*)$, and it is important to make this distinction. Instead of estimating a parameter we predict an observation. 
}
or BLUP. That is, to find an unbiased linear estimator which has minimum variance error. In mathematical terms, consider the (linear) statistic 
%
\begin{equation}
    T(Y)=\displaystyle \sum_{i=1}^n \lambda_i Y_i + \lambda_0= \begin{bmatrix}
    \lambda_0 & \bm{\lambda}^\top
\end{bmatrix}\begin{bmatrix}
    1 \\ Y
\end{bmatrix}
\end{equation}


%
where $Y_i=Y(\bm{x}_i)$. Searching the BLUP implies the following optimisation problem: 
\begin{mini*}|s|
{\bm{\lambda} \in \mathbb{R}^n}{\mathbb{V}\big[\sum_{i=1}^n \lambda_i Y_i + \lambda_0 - Z(\bm{x}_*)\big]  = \\
= \Scale[0.85]{\hspace{-0.1cm} \mathbb{V}\big[\sum_{i=1}^n \lambda_i Y_i\big]+\mathbb{V}\big[Z(\bm{x}_*)\big]  
-2\mathrm{Cov}\big(\sum_{i=1}^n \lambda_i Y_i + \lambda_0, \ Z(\bm{x}_*)\big)}}{}{}
\addConstraint{\Scale[0.85]{\hspace{-1.5cm} \mathbb{E}[\sum_{i=1}^n \lambda_i Y_i+  \lambda_0]=m(\bm{x}^*)}}
\end{mini*}
%
%
%
%

Note that, in general, to have an unbiased estimator does not guarantee the best estimator in terms of mean squared error. See, for example, \citet{hardy2003} for counterexamples. 

%
%
Some authors assume that the mean function is zero or subtract it from each observed value to obtain a non-restricted optimization problem. Here, I do not take this approach; instead, I use the more general method of Lagrange multipliers, which will later be useful for describing OK. First, let us express the three terms of the objective function in vector notation:

\vspace{-5mm}

\begin{align}
    \mathbb{V}[\sum_i^n \lambda_i Y_i]&= \sum_{i,j=1}^n \lambda_i \lambda_j \mathrm{Cov}(Y_i,Y_j)  \\
    &=\sum_{i,j=1}^n \lambda_i \lambda_j \mathrm{Cov}(Z(\bm{x}_i)+\varepsilon_i, \ Z(\bm{x}_j)+\varepsilon_j) \notag \\ \notag
    & = \bm{\lambda}^\top \Sigma \bm{\lambda} + \bm{\lambda}^\top \sigma^2 \bm{\lambda}=\bm{\lambda}^\top(\Sigma+\sigma^2I)\bm{\lambda}.
\end{align}
where $\Sigma=\begin{pmatrix} k(\bm{x}_i,\bm{x}_j)
    \end{pmatrix}_{i,j=1,2,...,n}$ and $\sigma^2=\mathbb{V}[\varepsilon]$.
%



\begin{align}
    \mathbb{V}\big[Z(\bm{x}_*)\big]:=\sigma^2_*
\end{align}

%
\begin{align}
     &\mathrm{Cov}(\sum_{i=1}^n \lambda_i Y_i + \lambda_0, Z(\bm{x}_*))  \\
     &=\sum_{i=1}^n \lambda_i \mathrm{Cov}(Z(\bm{x_i})+\varepsilon_i, Z(\bm{x}_*)) = \bm{\lambda}^\top \bm{k}_* \notag
\end{align}

where $\bm{k}_*= \big(k(\bm{x}_1,\bm{x}_*), k(\bm{x}_2,\bm{x}_*),...,k(\bm{x}_n,\bm{x}_*)\big)^\top$.
Thus, the objective function is 
\begin{equation}
    f(\begin{bmatrix}
    \lambda_0 & \bm{\lambda}
\end{bmatrix}^\top)= \bm{\lambda}^\top(\Sigma+\sigma^2I)\bm{\lambda} + \sigma^2_* -2 \bm{\lambda}^\top \bm{k}_*.
\end{equation}


On the other hand the restriction results in,
\begin{align}
    \hspace{-0.5cm} \mathbb{E}[\sum_{i=1}^n \lambda_i Y_i + \lambda_0] &= \sum_{i=1}^n \lambda_i \mathbb{E}[Y_i] + \lambda_0=\sum_{i=1}^n \lambda_i m(\bm{x}_i) + \lambda_0 \\ \notag
    &=\begin{bmatrix}
    \lambda_0 & \bm{\lambda}^\top 
\end{bmatrix}
\begin{bmatrix}
    1 \\
    \bm{m}
\end{bmatrix} \triangleq g(\begin{bmatrix}
    \lambda_0 \\ \bm{\lambda}
\end{bmatrix})
\end{align}
where $\bm{m}= \big(m(\bm{x}_1),m(\bm{x}_2),...,m(\bm{x}_n)\big)^\top$. Notice that both functions, $f$ and $g$, are of class $C^\infty$ since they are quadratic and linear functions of the variables, respectively. In addition, $f$ is a convex function (the Hessian matrix is positive semi-definite) and $g=m(\bm{x}^*)$ defines a convex set, therefore we have a convex optimisation problem.

%

Applying the method of Lagrange multipliers, we obtain the necessary conditions for a solution, which are given by the following system of equations:
\begin{equation}
  \begin{cases}
    \nabla f= \mu \nabla g,\ \ \mu\in \mathbb{R} \\
    g(\bm{\lambda})=m_*
  \end{cases}
\end{equation}
along with any other points, if any, satisfying the constraint $g(\bm{\lambda})=m_*$, and such that $ \nabla g$ vanishes. Note that $m_* \triangleq m(\bm{x}_*)$. 
%
%
Using the denominator convention for matrix derivatives we get that 
\begin{equation}
  \begin{cases}\label{eq:ksys}
  \begin{bmatrix}
    0 \\
      2(\Sigma+ \sigma^2I)\bm{\lambda}-2\bm{k}_*
  \end{bmatrix} = \mu \begin{bmatrix}
      1 \\ \bm{m} 
  \end{bmatrix} \longrightarrow  \mu = 0
    \\
\begin{bmatrix}
    \lambda_0 & \bm{\lambda}^\top 
\end{bmatrix}
\begin{bmatrix}
    1 \\
    \bm{m}
\end{bmatrix} = m_* \longrightarrow \lambda_0=(m_* -\sum_{i=1}^n \lambda_i m_i )
  \end{cases}
\end{equation}
In this case, $\mu$ is easily found to be 0 and $\lambda_0$ can be computed in terms of the other values of $\bm{\lambda}$. A Lagrange multiplier of 0 indicates that the constraint corresponding to that multiplier does not affect the optimisation process, at the point where it's evaluated. It means that either the constraint is inactive at that point or it has no impact on the objective function in that specific context, the former being the case here.
Finally, solving for $\bm{\lambda}$ we get the solution:
\begin{align}
     2(\Sigma+ \sigma^2I)\bm{\lambda}-2\bm{k}_*&=0 \notag \\ \notag
     (\Sigma+ \sigma^2I)\bm{\lambda}&=\bm{k}_* \\
     \bm{\lambda}&= (\Sigma+ \sigma^2I)^{-1}\bm{k}_*
\end{align}
Note that the matrix $\Sigma$ is invertible whenever the random variables $\{Z(\bm{x_i})\}_{i=1,2,...,n}$ are not degenerated. Anyway, the ``inflation term'' $\sigma^2I$ makes $(\Sigma+ \sigma^2I)$ to be invertible for $\sigma^2 \neq 0$.
Thus, the weights of the linear estimator are:

$\begin{cases}
    \hat{\bm{\lambda}} =  (\Sigma+ \sigma^2I)^{-1}\bm{k}_* \\
    \hat{\lambda}_0=(m_* -\sum_{i=1}^n \lambda_i m_i )
\end{cases}$

A measure of uncertainty of the prediction is a very desirable aspect and is indeed considered in the Kriging literature. The variance of the estimator is 
%
\begin{align}
    \hspace{-0.5cm} \mathbb{V}(T(Y))&=\mathbb{V}(\bm{\lambda}^\top Y+\lambda_0)=\mathbb{V}(\bm{\lambda}^\top Y)= \lambda^\top \mathbb{V}(Y) \lambda  \notag \\ \notag
    &= \bm{k}_*^\top (\Sigma+ \sigma^2I)^{-1}(\Sigma+ \sigma^2I) (\Sigma+ \sigma^2I)^{-1}\bm{k}_*  \\
    &=\bm{k}_*^\top(\Sigma+ \sigma^2I)^{-1}\bm{k}_*
\end{align}


%
and the variance of the estimation error is (also known as mean squared error):
%
\begin{align}
    \mathbb{V}\big[T(Y)-Z(\bm{x}_*)\big]&=f(\begin{bmatrix}
        \hat{\lambda}_0 & \hat{\bm{\lambda}})
    \end{bmatrix}^\top) = (...) \notag \\ 
    &= \sigma^2_* - \bm{k}_*^\top(\Sigma+\sigma^2I)^{-1}\bm{k}_*  \\
    &=  \mathbb{V}[Z(\bm{x}_*)]-\mathbb{V}\big[T(Y)\big] \notag
\end{align}


%
Observe that the variance does not depend on the observed values $Y$, but on covariances. It is composed by the variance of the random field at the estimation point minus a reduction term due to the best linear approximation. 

As said, the different variants of Kriging differ by the assumption made on the mean structure. Let's present them.

\subsection{Simple Kriging}


SK corresponds to the previous problem when the mean function $m$ is considered known and there is no noise $(\sigma^2=0)$.
%
Matheron did not specifically use the term ``Simple Kriging'' in his early works, but started to use it in his landmark work
\citet[Sec. 4.3.3]{matheron1971}, with simply referring to the knowledge of the mean function.
There is some variability in the literature about the assumptions for the mean structure in SK, but all agree that is the case where the mean function is known.  For example, \citet[Sec. 1.5]{stein1999} considers that SK refers to $m(\bm{x})=0$,  \citet{dalmau2017,chiles2018, oliver2007} that $m(\bm{x})=\text{const.}$, whereas \citet{cressie2015} considers that $m$ can be any known function. The last case is the most general, and the one considered here.
 
Following the results of the previous section, when there is no noise the solution simplifies to:
\begin{equation}
\begin{cases}
    \hat{\bm{\lambda}} =  \Sigma^{-1}\bm{k}_* \\
    \hat{\lambda}_0=(m_* -\sum_{i=1}^n \lambda_i m_i )
\end{cases}
\end{equation}

and the variance of the prediction error simplifies to:
%
\begin{equation}
    \mathbb{V}\big[T(Y)-Z(\bm{x}_*)\big] =\sigma^2_* - \bm{k}_*^\top\Sigma^{-1}\bm{k}_*
\end{equation}


%
With a bit of algebra the estimation of $Z(\bm{x}_*)$ results in:
%
\begin{align}
    \hat{Z}(\bm{x}_*)=\displaystyle \sum_{i=1}^n \hat{\lambda}_i Y_i + \hat{\lambda}_0 &= \hat{\bm{\lambda}}^\top Y + (m_* -\hat{\bm{\lambda}}^\top \bm{m}) \notag \\ \notag 
    &=m_* + \hat{\bm{\lambda}}^\top(Y-\bm{m}) \\
    &=m_* + \bm{k}_*^\top \Sigma^{-1}(Y-\bm{m})
\end{align}


%
As we see, the estimator adds up to $m_*$, the mean function at $\bm{x}^*$, a `correction term' which is a weighted sum of the differences to the mean of the observations. 




 

 

The equivalence of the results to the case in which the mean is subtracted at the beginning (to work directly with a zero-mean version of the random field) is shown in Appendix~\ref{ap:zero-mean_sk}.
Subtly, this equivalence only occurs if we consider an estimator of the form $T(Y)= \sum_{i=1}^n \lambda_i Y_i + \lambda_0$, not of the form  $T(Y)= \sum_{i=1}^n \lambda_i Y_i$, without the term $\lambda_0$. In that case, the estimator would have been different, which is the case treated in OK with the additional supposition that the mean function is constant and unknown. 
See `\textit{Homogeneous and Heterogeneous Linear Predictor}' paragraph in \citet[pg.178]{cressie2015} for an overview of the classes of linear estimators we can choose, from smallest to largest MSE.



Finally, a last remark. If the mean function is a known constant,   $m(\bm{x})=c$, then $\hat{\lambda}_0$ simplifies to $\hat{\lambda}_0= c(1 -\sum_{i=1}^n \lambda_i )$. Some authors \citep{chiles2018, oliver2007} assume that the mean is a known constant and directly introduce SK as an estimator of the form  

%
\begin{equation}
    T(Y)=\displaystyle \sum_{i=1}^n \lambda_i Y_i + (1-\sum_{i=1}^n \lambda_i)c
\end{equation}


%
without explicitly stating that this expression originates from considering a linear estimator of the form $T(Y)= \sum_{i=1}^n \lambda_i Y_i + \lambda_0$, and imposing the unbiasedness constraint. 

\subsection{Ordinary Kriging}

OK refers to the case of an unknown constant mean, $m(\bm{x})=c$, and no noise, $\sigma^2=0$. In the literature without providing a justification, an estimator of the form $T(Y) = \sum_{i=1}^n \lambda_i Y_i$, with no independent term $\lambda_0$ as in SK, is considered. We may soon see a reason. Putting all this information together in Eq. \eqref{eq:ksys}, the Kriging system reduces to:
%
\begin{equation}
\begin{cases}
   2\Sigma\bm{\lambda}-2\bm{k}_*= \mu  c \bm{1} \\
   \bm{\lambda}^\top  \bm{1} =1
\end{cases}
\end{equation}
%


where in the second equation $c$ cancels out. Is common in the literature to find this last constraint as chosen \citep{matheron1971, chiles2018, oliver2007}, but in fact, it arises naturally from imposing the unbiasedness constraint. Matheron \citep{matheron1971} calls it the `universal condition'.

If we look at the equations, we see that the system is linear in terms of $\bm{\lambda}$ and $\mu$. Thus, with a bit of algebra we can rearrange the terms as:
%
\[ 
  \begin{cases}
   \Sigma\bm{\lambda}+\bm{1} (-0.5 c \cdot \mu )= \bm{k}_* \\
  \bm{1}^\top \bm{\lambda}   =1
  \end{cases}
\]
and write it in matrix form as:
%
\begin{align}
 \left[\begin{array}{cc}\label{eqq:ks2}
\Sigma & \mathbf{1} \\
\mathbf{1}^\top & 0
\end{array}\right]   \left[\begin{array}{c}
\bm{\lambda} \\
\tilde{\mu}
\end{array}\right]=\left[\begin{array}{c}
 \bm{k}_* \\v%
1
\end{array}\right]
\end{align}
with $\tilde{\mu}=-0.5 c \mu $. The absorption of $c$ in the Lagrange multiplier is a clever step, and may be the reason why, in OK, a purely linear estimator is considered. By absorbing $c$ in the Lagrange coefficient, we can continue as if $m$ were known, avoiding the necessity of estimation. Indeed, for the prediction of the value $Z_*$ together with his variance, only the estimate of $\bm{\lambda}$ is needed.

Equation \eqref{eqq:ks2} is known in the literature as the \textbf{Kriging System} (KS) and is typically not  developed further.
Nevertheless, under some inverse assumption, it exists a closed-form solution which can be found by using block-Gaussian elimination and the Schur complement.

In general, consider a matrix
%
%
$S = \scriptstyle \left(\begin{smallmatrix}  A & B \\ C & D \end{smallmatrix}\right)$
where the inverse of the block $A$ and the Schur complement of it, $D-CA^{-1}B$, exist. Then,
%
\begin{align}\label{eq:inv_lemma}
\hspace{-.75cm} S^{-1}= \left(\begin{array}{cc}
\scriptstyle A^{-1}+A^{-1} B\left(D-C A^{-1} B\right)^{-1} C A^{-1} & \scriptstyle -A^{-1} B\left(D-C A^{-1} B\right)^{-1} \\ \scriptstyle
-\left(D-C A^{-1} B\right)^{-1} C A^{-1} & \scriptstyle \left(D-C A^{-1} B\right)^{-1}
\end{array}\right).   
\end{align}
See \citet[Proposition 3.9.7.]{bernstein2009} for more details.
If we apply the inversion preposition to the Kriging System with $A=\Sigma, \ B=\bm{1},\ C=\bm{1}^\top$, and $D=0$, after some rearrangement we get
%
\begin{equation}
\bm{\hat{\lambda}}=\Sigma^{-1}\Big(\bm{k}_*+ {\tilde{\mu}}\bm{1} \Big) , \quad {\tilde{\mu}}=\frac{\left(\bm{1}^\top \Sigma^{-1} \bm{k}_* -1\right)}{\bm{1}^\top \Sigma^{-1} \bm{1}}  
\end{equation}


%
Now the estimator of $\bm{\lambda}$, in comparison to SK, has an additional correction term ${\tilde{\mu}}$ which is added to $\bm{k}_*$. 
%
%
This solution appears in some texts without explanation or reference to its origin, such as in \citet[Eq. 5.12]{kleijnen2008} and \citet[Eq. 4]{kleijnen2017}.

By definition, the variance of the estimator error is
%
\begin{align}
    \mathbb{V}\big[T_{\text{OK}}(Y)-Z(\bm{x}_*)\big]&=\mathbb{V}\big[\hat{\bm{\lambda}}^\top Y- Z(\bm{x}_*)\big] \notag \\ \notag
    &= \sigma^2_*  + \hat{\bm{\lambda}}^\top \Sigma\hat{\bm{\lambda}}   -2\hat{\bm{\lambda}}^\top\bm{k}_*  = (...) \\
    &= \sigma^2_{SK} + \frac{(1-\bm{1}^\top \Sigma^{-1} \bm{k}_*)^2}{\bm{1}^\top \Sigma^{-1} \bm{1}} 
\end{align}

Hence the variance of OK is the variance of SK plus an extra term coming from the fact that we are using an estimator without (the extra degree of freedom term) $\lambda_0$, but which allows $m$ to be unknown.

In the literature, the variance is typically represented in the compact form \citep{chiles2018}:
%
\begin{equation}
\sigma^2_* - \bm{\hat{\lambda}}^\top \bm{k}_*-\tilde{\mu} 
\end{equation}
No explicit form for $\tilde{\mu}$ is given, since the OK system is usually not solved in a closed form. It is considered the last step, before bringing computers into play.

In the case of trying to find an estimator with independent term $\lambda_0$, we would get the solution of SK, for the specific case $m(\bm{x})=c$, which would depend on the value $c$. The `absorption trick' from Eq. \eqref{eqq:ks2} could not be done. Since $c$ is unknown, we don't have an estimator. A way to solve this is to plug in for $c$ a Generalised Least Squares (GLS) estimator for the mean value which is: $\hat{c}=\frac{1}{\bm{1}^\top \Sigma^{-1}\bm{1}} \bm{1}^\top\Sigma^{-1} Y$. The Appendix \ref{ap:glsOK} shows that this two step procedure is equivalent to OK.

Finally, in the Appendix \ref{ap:OK direct} an alternative way of solving the OK system without using the inversion preposition is shown. 


\subsection{Universal Kriging}

UK is a generalisation of OK by considering that the mean function is now of the form 
\begin{equation}
m(\bm{x})=\sum_{l=1}^p f_l(\bm{x}){\beta}_l= \bm{f}^\top \bm{\beta}
\end{equation}


%
where:

\begin{itemize}
    \item $\bm{f}$ are a collection of $p$ known functions, for instance, a truncation of a polynomial base.
    \item $\bm{\beta}$ is an unknown vector of parameters.
\end{itemize}

The unbiasedness constrain is expressed as:
%
\begin{align*}
    \mathbb{E}[T(Y)]&=\mathbb{E}[\sum_{i=1}^n \lambda_i Y_i] = \sum_{i=1}^n \lambda_i \mathbb{E}[Y_i]  \\
    &=\sum_{i=1}^n \lambda_i \sum_{l=0}^p f_l(\bm{x}_i)\bm{\beta}_l=\underbrace{\bm{\lambda}^\top M\bm{\beta}}_{g(\bm{\lambda})}
\end{align*}


%
where
%
\begin{align*}
    M_{ij}=f_{j}(\bm{x}_i), \ i=1,...,n, \ \ j=1,...,p \\
    \bm{f}(\bm{x}_*)=\big(f_{1}(\bm{x}_*),...,f_{p}(\bm{x}_*)\big).
\end{align*}

Thus, the unbiasedness constrain can be written as
\begin{equation}
(\bm{\lambda}^\top M -\bm{f}(\bm{x}_*)^\top)\bm{\beta}=0   
\end{equation}


%
which is satisfied if (we can not cancel out $\bm{\beta}$ since is a vector):
\begin{enumerate}
    \item $\bm{\lambda}^\top M -\bm{f}(\bm{x}_*)^\top=0$ or,
    \item $\bm{\beta}=\bm{0}$ or,
    \item it exists a linear combination such that the sum-product is zero.
\end{enumerate}

The first option is considered in UK, because it will allow to use the same `trick' as in OK, to avoid the necessity to estimate $\bm{\beta}$ by absorbing the parameters in the Lagrange multipliers. Cressie \citep{cressie2015} called this relaxation (restriction 1) \textit{uniformly unbiased}, whereas to consider just $(\bm{\lambda}^\top M -\bm{f}(\bm{x}_*)^\top)\bm{\beta}=0$ weakly unbiased.

The Lagrange equation, from which we derive the UK system is:
%
\[ 
  \begin{cases}
    \nabla f= \mu \nabla g\\
    g(\bm{\lambda})=\bm{f}(\bm{x}_*)^\top \bm{\beta}
  \end{cases} 
\]
which results in:
\[
\begin{cases}
   2\Sigma\bm{\lambda}-2\bm{k}_*= \mu \cdot  M \bm{\beta}  \\
   \bm{\lambda}^\top M\bm{\beta}=\bm{f}(\bm{x}_*)^\top \bm{\beta} \underbrace{\longrightarrow}_{\text{UK}} \bm{\lambda}^\top M=\bm{f}(\bm{x}_*)^\top.
\end{cases}
\]
Rearranging,
\begin{equation}
  \begin{cases}
   \Sigma\bm{\lambda}+ M \cdot (-0.5\mu \bm{\beta})=\bm{k}_* \\
 M^\top \bm{\lambda} =\bm{f}(\bm{x}_*)
  \end{cases}    
\end{equation}


By re-defining $\Tilde{\bm{\mu}}=(-0.5\mu \bm{\beta})$ we get a system of $p$ Lagrange multipliers which is known in the literature as the UK system \citep{oliver2007}:
\begin{align}
 \left[\begin{array}{cc}
\Sigma & M \\
M^\top & 0
\end{array}\right]   \left[\begin{array}{c}
\bm{\lambda} \\
\tilde{\bm{\mu}}
\end{array}\right]=\left[\begin{array}{c}
\bm{k}_{*} \\
\bm{m}(\bm{x}_*)
\end{array}\right]
\end{align}\label{eq:ks3}
If $M$ and $\Sigma$ are of full rank, following the same approach as the one for OK (eq. \eqref{eq:inv_lemma}), we can get the following solution,
\begin{align}
\hat{\boldsymbol{\lambda}}=\left\{\Sigma^{-1}-\Sigma^{-1} M\left(M^\top \Sigma^{-1} M\right)^{-1} M^\top \Sigma^{-1}\right\} \bm{k}_* + \notag \\ 
\Sigma^{-1} M\left(M^\top \Sigma^{-1} M\right)^{-1} \bm{m}\left(\bm{x}_*\right)    
\end{align}


%
and 
\begin{equation}
T(Y)=\hat{Z}_*=\hat{\bm{\lambda}}^\top Y= \bm{f}(\bm{x}_*)^\top \hat{\bm{\beta}}    + \bm{k}_*^\top\Sigma^{-1}(Y-M\hat{\bm{\beta}})
\end{equation}
with $\hat{\bm{\beta}}=\left(M^T \Sigma^{-1} M\right)^{-1} M^T \Sigma^{-1}Y$ being the GLS estimator of $\hat{\bm{\beta}}$.

Two observations are important here:
\begin{itemize}
    \item[a)] If we let $p=1$ and $f(\bm{x})=1$ then $m(\bm{x})=\beta$, and we get the OK solution.
    \item[b)] If we apply SK to $Y-\bm{m}$, acting as if the mean function (in the form of UK) were known, and then we plug in the GLS estimator of $\bm{\beta}$, we get the same result (see \citet[Sec. 3.4.5]{cressie1990}).
\end{itemize}

The variance of the estimator error results in:
\begin{equation}
\sigma^2_{UK}=\sigma^2_{SK}+\gamma(M^\top\Sigma^{-1}M)^{-1}\gamma, 
\end{equation}


%
with  $\gamma= \bm{m}(\bm{x}_*)^\top -M^\top\Sigma^{-1}\bm{k}_*$. We may ask what happens if $Z_*$ is exactly one of the points $Z_i$. Since SK, OK and UK consider no noise, $\sigma^2=0$, the estimators interpolate the data. Thus, Kriging is an exact interpolator. This can be seen by noticing that when $Z_*=Z_i$, for some $i$, the best linear estimator is $\lambda=(0,\ldots,1,0,\ldots,0)^\top Y$, with the one in the $i-th$ position. It is not difficult to see, that when plugging this $\lambda$ in the error variance formula, it results that the variance is 0. On the other hand, if there is noise and we updated the Kriging formulae to include it, then we would get a smoothing method, now with variance greater than 0 when we predict over the observed values.

In Table \ref{t:k_summary} a summary of the Kriging methods can be found, which is discussed in the next section.

\section{The bridge}\label{sec:stairs}

In linear regression there are three main approaches to estimate the parameters of the regression hyperplane. They are: 

\begin{enumerate}
    \item[a)] Minimising the square of the errors. In other words, the ordinary least square procedure. There are no probability distribution assumptions nor any stochastic interpretation.

    \item[b)] Finding the BLUE. Again there are no probability distribution assumptions, but now we are using the concept of random variable and his first two moments. No other moments are used.

    \item[c)] Likelihood method. A probability distribution for the errors is assumed, typically the normal. 
\end{enumerate}

Once the regression is made, the regression hyperplane can be used to predict, which is the aim of Kriging. Consider that we have the following model:
%
\begin{equation}\label{eq:reg}
    y_i=m(\bm{x_i})+  \Tilde{Z}(\bm{x_i}) + \varepsilon_i, \ \ i=1,2,...,n 
\end{equation}


where


%
\begin{itemize}
    \item $m$ is any function, which can be interpreted as a mean function.
    \item $\Tilde{Z}$ a zero-mean random field.
    
    \item $\varepsilon$ are an i.i.d. collection of random variables, which can be interpreted as white noise.
\end{itemize}

Note that $\varepsilon$ can be absorbed in $\Tilde{Z}$ but I have considered it separately, for the sake of comparison with linear regression. 
It results that in the previous regression equation, depending on the mean hypothesis and the estimation procedure we have different predictors, which are represented in Table \ref{t:k_summary}. The first column represents the different mean hypotheses for the regression equation \eqref{eq:reg}. The second column represents the procedure a), which is to apply ordinary LS in Eq. \eqref{eq:reg} (this ignores the term $\Tilde{Z}$). The third column represents the procedure b), applied twice, a BLUP for the predictor (supposing the mean is known) and a BLUE for the mean, which ends up to be equivalent to the different variants of Kriging. Finally, the last column would correspond to procedure c), which is the GPR setting and is detailed in the next section. 

\begin{table*}[h!]
\caption{This table shows the results of the estimation procedures a), b) and c) stated in Sec. \ref{sec:stairs} applied to the Kriging problem of estimating an unobserved value ${y}_*$. Interesting connections appear between, LS, GLS, Kriging, and GPR. Since LS is not considering any probabilistic interpretation it is ignoring any correlation structure. Then, Kriging generalised the LS procedure by taking into account the correlation structure. The next step GPR, takes into account not only the correlation structure, but the whole probability distribution.}\label{t:k_summary}
\resizebox{\linewidth}{!}{
\begin{tabular}{@{}|l|l|l|l|@{}}
\hline
Mean hypothesis &   
Least Squares    & 
\begin{tabular}[c]{@{}l@{}}  BLUE for mean + BLUP for $Z_*$  \\  equivalent to Kriging \end{tabular} &  
\begin{tabular}[c]{@{}l@{}} ML for mean  + \\  cond.  dist. for $Z_*$   \end{tabular}    \\ \hline
$m$ known & $\hat{y}_*=m(\bm{x})$ & 
\begin{tabular}[c]{@{}l@{}} $ \hat{y}_*=m_* + \Sigma^{-1}\bm{k}_*(Y-\bm{m})$  \\  SIMPLE KRIGING \end{tabular}  &  
\multirow{3}{*}{GPR}\\ \cline{1-3}
\begin{tabular}[c]{@{}l@{}} $m(\bm{x})=c$  \\  c unknown \end{tabular}  &   
\begin{tabular}[c]{@{}l@{}} $\hat{y}_*=\hat{a}^{LS}$  \\  $\hat{a}^{LS} =(\bm{1}^\top \bm{1})^{-1}\bm{1}Y=\bar{Y}$ \end{tabular}      & 
\begin{tabular}[c]{@{}l@{}} {} \\ $\hat{y}_*=\hat{c}^{GLS}+\bm{k}_*^\top \Sigma^{-1}(Y-\hat{c}^{\scriptscriptstyle GLS}\bm{1})$  \\  $\hat{c}^{\scriptscriptstyle GLS}=\frac{1}{\bm{1}^\top \Sigma^{-1}\bm{1}} \bm{1}^\top\Sigma^{-1} Y$ \\ ORDINARY KRIGING \\ {} \end{tabular} 
&                      \\ \cline{1-3}
\begin{tabular}[c]{@{}l@{}} $m(\bm{x})=\displaystyle\sum_{l=1}^p f_i(\bm{x})\beta_l$  \\  $\bm{\beta}$ unknown \end{tabular} &     
 \begin{tabular}[c]{@{}l@{}} $\hat{y}_*=\bm{f}(\bm{x}_*)^\top\hat{\beta}^{LS}$  \\  $\hat{\beta}^{LS} =(M^\top M)^{-1}M^\top Y$ \end{tabular}  &  
\begin{tabular}[c]{@{}l@{}} {} \\ $\hat{y}_*=\bm{f}(\bm{x}_*)^\top \hat{\bm{\beta}}^{\scriptscriptstyle GLS} + \bm{k}_*^\top\Sigma^{-1}(Y-M\hat{\bm{\beta}}^{\scriptscriptstyle GLS})$  \\ $\hat{\bm{\beta}}^{GLS}=\left(M^T \Sigma^{-1} M\right)^{-1} M^T \Sigma^{-1}Y$ \\  UNIVERSAL KRIGING \\ {}\end{tabular}  
&   \\ \hline
\end{tabular}
}

\end{table*}
\section{Mathematical formulation of GPR}\label{sec:gpr}


In this section, the mathematical formulation of GPR, which is necessary for the comparison with Kriging, is briefly introduced. For a comprehensive review, see \citet{rasmussen2006}.

Consider that we have some observations from a random field, where now the additional supposition of Gaussianity is made. That it, any sample of the random field, $(Z(\bm{x}_1), ..., Z(\bm{x}_n))$, is multivariate Gaussian distributed. Consider that the mean function, $m$, and the autocovariance function, $k$, is known and that we possibly have some additive i.i.d. Gaussian noise,
%
\begin{equation}
    (Y(\bm{x}_1), \ldots, Y(\bm{x}_n)) \sim    N(m(X),\ k(X,X)+\sigma^2 I)
\end{equation}


where $X$ is an $n \times d$ matrix filled with the location values by rows, sometimes called design matrix. To evaluate $m$ or $k$ at $X$ means, in this notation, to form a vector or matrix, respectively, with the elementwise evaluations of $X$.

We are interested in predicting the random field in a new location $\bm{x}_*$. Because of the Gaussianity assumption the joint distribution of the sample and $Z_*$ will be also Gaussian,
\begin{equation}
\Scale[0.9]{\left[\begin{array}{c}
{Y} \\
Z_*
\end{array}\right] \sim \mathcal{N}\left(\left[\begin{array}{c} m(X) \\ m(\bm{x}_*)\end{array}\right], \left[\begin{array}{cc}
k(X, X)+\sigma^2 I & k\left(X, \bm{x}_*\right) \\
k\left(\bm{x}_*, X\right) & k\left(\bm{x}_*, \bm{x}_*\right)
\end{array}\right]\right)}
\end{equation}

\vspace{-3mm}

To predict we are interested in the conditional distribution of $Z_* \mid Y=\bm{y}$. This predictive distribution is also Gaussian and can be explicitly obtained \citep[Appendix 2]{rasmussen2006} resulting in ($\sigma^2$ is assumed to be 0 as in Kriging):
%
\begin{equation}\label{eq:gpr_pred}
\begin{aligned}
& z_* \mid \bm{y} \sim \mathcal{N}\left(m(\bm{x}_*) + k\left(\bm{x}_*, X\right) k(X, X)^{-1} (\mathbf{y}-m(X)),\right. \\
&\left.k\left(\bm{x}_*, \bm{x}_*\right)-k\left(\bm{x}_*, X\right) k(X, X)^{-1} k\left(X, \bm{x}_*\right)\right) .
\end{aligned}
\end{equation}


%
In attention to the previous sections if we rename $m_*=m(\bm{x_*})$, $\bm{m}=m(X)$, $\sigma_*^2=k(\bm{x}_*, \bm{x}_*)$, $\bm{k}_*= k(X, X_*)$, and $\Sigma=k(X, X)$ it results that the Gaussian random variable has as expected value the SK predictor and variance the SK prediction error. Thus, if we take the Maximum a Posterior (MAP) as a predictor of $Z_*$, it will correspond to the SK one. 
This is just a possible choice for the punctual estimator, where now we have the knowledge of the whole probability distribution.

One may wonder if we are using somehow a Bayesian approach. The answer is yes in some sense, since we are predicting $Z_*$ by using the available information, nevertheless, since the joint distribution of $Y$ and $Z_*$ is directly available, the Bayesian formulae is not useful.

\subsection{GPR with unknown mean}

Consider now that the mean is unknown and has the same form as in UK, $m(\bm{x})=\sum_{l=1}^p f_l(\bm{x})\beta_l$, where $\bm{\beta}$ should be inferred from the data. Recall that when  $p=1$, $f(\bm{x})=1$, the mean function is constant, as in OK.
In the GPR framework, we can assign a prior over the weights $\bm{\beta} \sim \mathcal{N}(\bm{b}, B)$. Using that the mean can be written apart as $Z(\bm{x})=f(\bm{x}) ^\top \bm{\beta}+  \mathcal{N}(0, k(\bm{x},\bm{x}))$, it is straightforward to see that the vector $\bm{z}=Z(X)=(Z(\bm{x}_1), ..., Z(\bm{x}_n))^\top$ is also Gaussian with 
%
\begin{align*}
    \mathbb{E}[\bm{z}]&=\mathbb{E}[M \bm{\beta}]=M\mathbb{E}[\bm{\beta}]=M\bm{b} \\
    \mathbb{V}[\bm{z}]&= \mathbb{V}[M \bm{\beta} + \mathcal{N}(0, k(X,X))] \\ 
    &=M^\top \mathbb{V}[\bm{\beta}] M + \mathbb{V}[\mathcal{N}(0, k(X,X))] \\ 
    &=M^\top B M + k(X,X)
\end{align*}


%
%
Now we can proceed in the same way as in Sec. \ref{sec:gpr}, taking the joint distribution of $[Y \ Z_*]^\top$ and then taking the conditional distribution of $Z_* \mid \bm{y}$. After simplification and taking the limit $B^{-1} \to 0$, to make the prior over $\bm{\beta}$ non-informative, we would obtain a normal predictive distribution with \citep[Eq. 2.42 ]{rasmussen2006}:
\begin{align}
     \mathbb{E}[z_* \mid \bm{y} ] &= \bm{k}_*^\top\Sigma^{-1}(\bm{y}-M\hat{\bm{\beta}}^\text{\tiny GLS})+\bm{m}(\bm{x}_*)^\top \hat{\bm{\beta}}^\text{\tiny GLS} \\
     \mathbb{V}[z_* \mid \bm{y} ]&=\sigma^2_{SK}+\gamma(M^\top\Sigma^{-1}M)^{-1}\gamma
\end{align}
with $\gamma= \bm{m}(\bm{x}_*)^\top -M^\top\Sigma^{-1}\bm{k}_*$.

If we use again the MAP to make a prediction it will coincide with the UK predictor and the variance of the MAP coincides with the prediction error in UK.

Notice that the predictive distribution is independent of the prior expected value $\bm{b}$ and that the GLS estimator also appears here. 
In fact, it is easy to check that the non-informative predictive distribution is equivalent to considering a likelihood estimator for $\beta$ in the GP setting, which would have given $\beta^{\scriptscriptstyle GLS}$, and plugging it in the conditional distribution for $Z_*$ (eq. \eqref{eq:gpr_pred}). Thus, the GPR method with an unknown mean is equivalent to the likelihood method + the conditional distribution for $Z_*$, as stated in Table \ref{t:k_summary}.

\section{Connections of the methods}\label{sec:5}

The relation between Kriging and GPR is of the same kind as the one in classical linear regression between the BLUE method and the likelihood method. GPR is a further step in the initial assumption about the problem, where not only the mean and autocovariance functions are used (the first two moments), but the entire probability law, which is specified as a GP.
However, compared to ordinary linear regression, Kriging and GPR have the additional objective of predicting the random variable of interest $Z_*$, using the correlation structure between the observations and the variable, and not only the regression hyperplane. The former is mean based whereas Kriging and the GPR MAP improve this estimation by adding a weighted sum of the differences between the observations and the mean (residuals). 
Kriging solves this problem by using a BLUP, whereas GPR uses a predictive conditional distribution.

When the mean function is known, it results that the SK estimator is the MAP of the GPR predictive distribution. In addition, the prediction error of SK coincides with the variance of the GPR predictive distribution.

On the other hand, when the mean is unknown, the OK and UK estimate $Z_*$ by a clever set up resulting in the so-called Kriging system, which hides in the Lagrange multipliers the mean value parameters. GPR tackles this problem by using a likelihood estimator for the mean value parameters, or by using a prior over $\bm{\beta}$ and tending the precision matrix of them to $\bm{0}$, both giving the same result.
Surprisingly or not, again the MAP of the GPR predictive distribution is the OK/UK estimators. In addition, the prediction error of OK/UK coincides with the variance of the GPR predictive distribution. Of course, GPR formalism provides not only a specific estimator but also the probability distribution of all the possible values.

%


In summary, the techniques use different terminology and approaches, reflecting both philosophical and technical differences. However, they are closely related: GPR is the natural extension of Kriging to 1) the likelihood-based framework and 2) the predictive conditional distribution under the Gaussianity assumption.




\subsection{Unknown kernel}

Along the work, the kernel was considered to be known or given. In the usual case, where the kernel is estimated from data, there are additional differences between Kriging and GPR \citep{demystifying2023}.
The kriging method fits the data through a two-step procedure. First, the observed data are transformed into the experimental variogram, which can be viewed as a noisy estimate of the underlying true variogram. Next, a parametric variogram is calibrated by minimising the least squares error between itself and the experimental variogram. In contrast, GPR directly optimises the kernel hyperparameters by maximising the marginal log-likelihood of the observations, without requiring an intermediate transformation. Because Kriging and GPR uses different optimisation objectives, the resulting optimal parameters---and consequently the predictions--- differs.

\section{Conclusion}\label{sec:concl}

The relationship between Kriging, a method with its origins in geostatistics, and GPR, a technique widely used in machine learning, has been explored. The historical overview provided context for the development of these methods.
Despite their distinct disciplinary roots, it has been shown that these methodologies share a common mathematical framework. 
%

The terminology used in Kriging largely differs from that used in GPR, including the mathematical formalism. 
An approach based on modern statistical terminology, with a unified and consistent presentation of both methods, has been provided. In addition, the Appendix~\ref{ap:k_terminology} explains common Kriging terminology in standard statistical terms.

Next, I have discussed how the GPR method can be viewed as a natural extension of Kriging. A comparative analysis of the three main types of Kriging—Simple Kriging, Ordinary Kriging, and Universal Kriging—have been conducted, highlighting both the similarities and differences with the corresponding GPR setup.
From a statistical point of view, GPR is the natural extension of the Kriging to the probabilistic framework under the Gaussianity assumption; the combination of the likelihood method for the estimation of the mean function with the predictive conditional distribution of the target variable.
%
In the supposition the variogram is unknown and is estimated from the data, the techniques have an additional difference. They optimise for different objectives, leading to related but unequal predictions.

In conclusion, this work aims to elucidate the relationship between the methods and facilitate the transfer of knowledge between disciplines.

\section*{Acknowledgements}

The author would like to acknowledge N. Cressie, for his 
notable clarity of exposition and his depth treatment of Kriging topics. Also to M. L. Stein, J. P. Chilès, P. Delfiner, C. E. Rasmussen, and C. K. I. Williams for their excellent work related to the topics discussed in this article.


\newpage

\bibliography{references}

\newpage
\section*{Appendix}

\begin{appendix}



\section{Equivalence of SK with zero-mean simplification}\label{ap:zero-mean_sk}

A question we may have is whether we get the same solution if we subtract the mean at the beginning, to work directly with a simple version of a zero-mean random field. Consider 
%
$$ Y(\bm{x}_i)-m(\bm{x}_i)=\tilde{Z}(\bm{x}_i)+\varepsilon_i, \ i=1,...,n$$


where $\tilde{Z}$ is a zero-mean random field.
Then working with $\tilde{Y}=Y-\bm{m}$ as being the initial observations, and applying the previous solution to the particular case that now the random field is zero-mean, we get

$\begin{cases}
    \hat{\bm{\lambda}}^{\tilde{Z}} =  \Sigma^{-1}\bm{k}_* \\
    \hat{\lambda}^{\tilde{Z}}_0=0
\end{cases}$
 
Then, the estimator of $\tilde{Z}$ would be
 $$\hat{\tilde{Z}}(\bm{x}_*) = 0 + \bm{k}_*^\top \Sigma^{-1}\tilde{Y}$$

 
and since $Z(\bm{x}_*)=m_*+\tilde{Z}(\bm{x}_*)$, the estimation results in:
  $$ \hat{Z}(\bm{x}_*)=m_* + \bm{k}_*^\top \Sigma^{-1}(Y-\bm{m})$$

which is the same as the SK solution. 

\section{Equivalency of OK with SK + GLS for the mean}\label{ap:glsOK}


The objective is to show that the OK predictor is equivalent to substituting the GLS estimate for the mean into the SK predictor. Starting from the OK predictor, a few algebraic manipulations allow to get the result,

\begin{align*}
    &T_{OK}(Y)=\hat{\bm{\lambda}}^\top Y=(\bm{k}_*^\top+\tilde{\mu}\bm{1}^\top)\Sigma^{-1}Y   \\
    &=\bm{k}_*^\top \Sigma^{-1}Y +\tilde{\mu}\bm{1}^\top\Sigma^{-1}Y  \\
    &=\bm{k}_*^\top \Sigma^{-1}Y +\frac{1-\bm{1}^\top\Sigma^{-1}\bm{k}_*}{\bm{1}^\top\Sigma^{-1}\bm{1}}\bm{1}^\top\Sigma^{-1}Y  \\
    &=\bm{k}_*^\top \Sigma^{-1}Y+(1-\bm{1}^\top\Sigma^{-1}\bm{k}_*) \hat{c}, \ \  \hat{c}=\frac{1}{\bm{1}^\top \Sigma^{-1}\bm{1}} \bm{1}^\top\Sigma^{-1} Y  \\
   &=\hat{c}+\bm{k}_*^\top \Sigma^{-1}(Y-\hat{c}\bm{1})
\end{align*}

which is the SK predictor with $m(x)=c$ changed by his GLS predictor. In \cite[Sec. 1.5]{stein1999} this is also stated in an alternative way and in a more general framework.

\section{Solving the OK system without inversion preposition}\label{ap:OK direct}

We can solve the OK system without the inversion preposition in a few steps by doing some clever matrix algebra. The OK system was:
\[ 
  \begin{cases}
   \Sigma\bm{\lambda}+\tilde{\mu}\bm{1}= \bm{k}_* \\
  \bm{1}^\top \bm{\lambda}   =1
  \end{cases}
\] 

We isolate $\lambda$ in the first equation:
\begin{align}
       \Sigma\bm{\lambda} + \tilde{\mu}\bm{1} = \bm{k}_* \longrightarrow \bm{\lambda}= \Sigma^{-1}\bm{k}_* - \Sigma^{-1}\tilde{\mu}\bm{1}  
\end{align}

which gives a solution depending on the Lagrange multiplier $\tilde{\mu}$ which is yet unknown. To find $\tilde{\mu}$ we multiply that equation by a vectors of ones (or equivalently we sum all the equations):
\begin{align*}
       \bm{1}^\top \bm{\lambda}&=
       \bm{1}^\top\Sigma^{-1}\bm{k}_* - \bm{1}^\top\Sigma^{-1}\tilde{\mu}\bm{1} \\
       1&= \bm{1}^\top\Sigma^{-1}\bm{k}_* - \bm{1}^\top\Sigma^{-1}\tilde{\mu}\bm{1}
\end{align*}
where we have substituted $ \bm{1}^\top \bm{\lambda}$ by one, using the unbiasedness constrain. The equation is scalar so we can isolate $\tilde{\mu}$
\begin{align*}
    \tilde{\mu} =-\frac{1-\bm{1}^\top\Sigma^{-1}\bm{k}_*}{\bm{1}^\top\Sigma^{-1}\bm{1}}
\end{align*}

Thus, the solution for $ \bm{\lambda}$ is
$$
\bm{\hat{\lambda}}=\Sigma^{-1}\Big(\bm{k}_*- {\tilde{\mu}}\bm{1} \Big) , \quad \text{with}  \quad {\tilde{\mu}}=\frac{-\left(1-\bm{1}^\top \Sigma^{-1} \bm{k}_*\right)}{\bm{1}^\top \Sigma^{-1} \bm{1}}.
$$
   
\section{Some common Kriging terminology explained (using statistics terminology)}\label{ap:k_terminology}

\begin{itemize}
    \item Nugget: the ``nugget effect'' means that the covariance of $Z(\bm{x})-Z(\bm{x}+\bm{h})$ does not tend to zero, when $\bm{h} \to \bm{0}$. Typically this is because there is noise in the observations, but it can also refer to ``short range'' variation in the random field \cite[Sec. 2.7]{matheron1971}, so the use of the term may be confusing as stated in \cite{YAKOWITZ1985}. N. Cressie defined it as the sum of both \cite{cressie1990}.


    \item Variogram: it refers to 
    \begin{align*}
        &2\gamma(\bm{x},\bm{x}')=\mathbb{V}[Z(\bm{x})-Z(\bm{x}')]= \\ &=\mathbb{V}[Z(\bm{x})] + \mathbb{V}[Z(\bm{x})] - 2\text{Cov}(Z(\bm{x}), Z(\bm{x}'))
    \end{align*}

    and is defined for practical purposes when trying to fit a model for the autocovariance structure. When the random field is stationary, isolating the covariance in the previous formula leads to the following relation: 
    $$\mathrm{Cov}(\tau)=\sigma_Z^2-\gamma(\tau).$$
    So, the variogram has an ``opposite'' interpretation to the covariance, the greater the covariance lesser the variogram and viceversa. 
    
    In the geostatistics literature, the Kriging estimator and the variance of the estimation error are usually expressed in terms of the variogram, but there is an equivalence in the formulae between that one and the one using covariances when the random field is second-order stationary. 
    
    In addition, Cressie argued in \cite{cressie1990} that the use of the variogram is preferable to the use of the autocovariance function. I quote a statement from his book: ``\textit{The cornerstone is the variogram, a parameter that in the past has been either unknown or unfashionable among statisticians}'' \cite[pg. 30]{cressie1990}.
    %
    %
    A more detailed treatment of this topic, is reserved for future work.
    


    
    \item Semivariogram: it refers to half of the variogram, $\gamma$. It is defined for practical purposes and for plotting. 

    \item Intrinsic (sense) stationary random field: it is a way to say that the mean function is constant and that the variance of $Z(\bm{x})-Z(\bm{x}')$ depends only on the difference $\bm{x}-\bm{x'}$ \cite[pg. 61]{cressie2015}. This intrinsic property does not imply wide-sense stationarity, a counterexample is the isotropic Brownian motion \cite[pg. 68]{cressie2015}, where the variogram depends on $x-x'$, but not the covariance. Even more, the variogram may exist when the autocovariance does not.
   
    
    \item Isotropic: when the random field is second order or weakly/wide-sense stationary and the covariance function, $C(\bm{x},\bm{x}' )$ is a function only of $||\bm{x} - \bm{x}'||$. \cite[Sec. 2.3.]{cressie2015}. Otherwise, it is called Anisotropic.

    The isotropic property can be seen as adding ``the norm'' to the intrinsic stationary condition.

\end{itemize}
\end{appendix}

\end{document}